# Effect of Brief Meditation Intervention on Attention: An ERP Investigation

Manvi Jain[1], C.M. Markan[2]

[1]Department of Cognitive Science, Dayalbagh Educational Institute, Agra

[2]Department of Physics and Computer Science, Dayalbagh Educational Institute, Agra

*Fast and efficient strategies for modulation of attention have been extensively studied recently. The present study has attempted to observe the effect of brief meditation practices on executive control of the attention system. The study recruits cognitive control by introducing conflict using Stroop task in twenty-six novice participants. Behavioral responses indicate a positive effect on response time and accuracy at Stroop task after ten minutes of meditation intervention. Neurophysiological findings suggest more efficient allocation of attentional resources. An increase in positive ERP components (P200, P300) and expected decrease in the inhibitory or negative component (N200) after intervention shows positive results. The findings suggest a positive impact of meditation intervention on attention even for brief periods in non-meditating population.*

*Index Terms*-- Attention, Conflict monitoring, Meditation intervention, P300, P200, N200

NOMENCLATURE

EEG  Electroencephalography (a method to record an electrogram of the brain's electrical activity)

ERP  Event-related potentials or Evoked response potentials are measured brain responses in terms of brain amplitude

P200  A visual P2 is a positive going electrical potential that peaks at about 200 milliseconds

P300  A positive deflection in voltage with a latency (delay between stimulus and response) of roughly 250 to 500 ms

N200  A negative-going wave that peaks 200 to 350 ms post-stimulus

## I. INTRODUCTION

Attention – the main component of meditation practice forms the basis of cognitive control in humans [1]-[4]. In the model of attention system, Posner and Petersen [5] proposed that the brain sources of attentional processes aggregate to form a system of three attentional sub-networks, each network carrying out functions of alerting, orienting, and executive control [6], [7].

Recent studies have focused on developing measures to modulate attention levels through cognitive and physical practices to overcome the expected limitations of human cognitive capacity [8],[9]. Broadly the following categories of cognitive-enhancement techniques have been introduced in the past: *Physical*, including aerobic exercise training [10]–[12] e.g., dance techniques [13] and brief strolls [14]; *Neurological*, including non-invasive brain-computer interface devices [15], invasive neural prosthetics [16] and transcranial stimulations, e.g., electrical, both direct and alternating [17] magnetic stimulations [18], focused ultrasound stimulations [19]; and *Cognitive*, including mindfulness meditation practices [20]–[23], multi-linguicism [24], [25], computerized cognitive training [26], [27].

The positive effect of cognitive exercises on attention such as meditation has been positively verified [28]–[32]. Kwak et al. [33] proved the positive effect of brief meditational retreats on attentional networks of the brain. It is therefore observed that any kind of intervention (physical, mental, emotional, etc.) might create an impact on attention, especially while performing extensive cognitive processes such as conflict monitoring [34]–[36].

Event-related potential (ERP) analysis suggest major changes in brain amplitudes as an effect of intervention. The ERP components reflect various cognitive processes. The major positive component appears around 250-350ms, reflects allocation of attentional resources of brain employed in a given attention task [37]. The component called as P300 component is majorly subject to change as a result of intervention. The early ERP components including negative N200 and positive P200 are linked to alerting (N2-alerting) and orienting processes [38] as given by the Posner-Peterson model of attention. In the present study, an attention network task (ANT) developed by J. Ridley Stroop (1935) called Stroop task [39], [42] was implemented to study the effect of a brief meditation intervention on attention. This was studied using ERP components in EEG signals, such as N200, P200 and P300.

Major studies implemented meditation to observe modulation of executive control implying direct impact of meditation practice on attention system. The present study hypothesizes positive effect of even brief meditation practice on brains of novice meditators. The implication of above hypothesis relates to temporary changes in attentional processes resulted by meditational practice, verified in further sections of the paper.

## II. MATERIALS AND METHODS

*1) Participants*

Twenty-six undergraduate students (mean age = 19.6, S.D.=1.5; 9 males and 17 females) participated in this study. The participants had no professional training in any meditational practice. All of them were right-handed and had normal or corrected-to-normal vision. A consent form was signed by each participant before starting the experiment in EEG testing.

*2) Apparatus and Stimulus*

An attention network task called Stroop word-colour task [39] was used as stimulus. The task design (Fig. 1) is composed of congruent stimuli consisting of the four colour words in English (green, yellow, purple, orange) written in the same colour as the word (e.g., word RED written in red colour ink). The incongruent stimuli consisted of another four colour words of English (green, yellow, purple, red) written in different combinations of colour-word and ink it is written in (e.g., word RED written in blue ink). In the incongruent condition, conflict is induced because of two differential stimuli simultaneously being presented to the subject. This test was designed using E-prime 3.0 (Psychology Software Tools, Sharpsburg, PA, USA).

The response buttons were programmed to be extreme left for incongruent stimulus and extreme right for congruent stimulus on a Chronos response box compatible with E-prime 3.0. The stimuli were presented for 1000ms followed by a fixation cross (a 'plus' sign) in black colour on a white screen for 500ms. During the 1000ms after presentation of stimulus, subjects were required to match the colour word with the colour ink and respond by pressing corresponding buttons – if the two features of the stimulus (colour and word) do not match (incongruent condition), the left button should be pressed, on the other hand, if they match (congruent condition), the right button should be pressed.

Before the first actual experiment, a practice test was performed by the subjects that included 20 stimuli with a feedback system. The feedback display reports correct and incorrect answers to the subject along with their accuracy between trials. The actual test consisted of total of 80 trials (40 congruent stimuli, 40 incongruent stimuli). After completion of the first part of Stroop test, subjects went through a brief intervention session discussed in the next section. After the intervention session, the subjects performed second part of Stroop test. The practice test was skipped in second test. During all these tasks and intervention practise, a 64-channel Electroencephalographic (EEG) system was used to measure their brain activity.

*3) Intervention Method*

After the first task, participants were asked to practice any kind of meditation for approximately ten minutes. As a derivative of meditation [40], a focused attention state was maintained by many participants. Majority participants selected to chant religious/meditative mantras as a mode of focusing attention. Consecutively, only the participants with mantra-based meditation intervention were selected for the analysis.

### III. DATA ACQUISITION

*1) Electrophysiological Recording*

The electrical activity of brain was recorded using a 64-channel EasyCap (Brain Products system) with sintered Ag/AgCI electrodes. The reference electrode was placed on the Cz, while ground was linked to the AFz. Sampling frequency was kept at 500 Hz samples and impedances were kept below 25 kΩ. The EEG was digitally low-pass filtered at 49 Hz and high-pass filtered at 1 Hz. Manually, filters were placed to reject artifacts of all types from the recording. Trials with ocular and muscular/movement artifacts were identified visually and excluded before averaging. The averaged epoch for ERP was 1500 ms, where the first 500 ms before stimulus presentation served as baseline for standardization purposes using baseline normalization technique (z-score method). Only segments with correct responses were averaged including around 30 trials for each condition (congruent and incongruent). All the pre-processing steps were performed under visual inspection with Brainstorm (Tadel et al. 2011), which is documented and freely available for download online under the GNU general public license [41]. The artifact-free averages of brain electrical activity across trials for all subjects were further analyzed using the same tool by applying ERP and source analysis techniques as discussed in further sections.

*2) Behavioural data*

Task designing tool ePrime 3.0 recorded Response time (in msec) for all trials separately. The data was extracted from ePrime 3.0 and exported for plotting and comparison. The reaction time data for congruent and incongruent trial conditions were separated for both first (pre-intervention) and second (post-intervention) test conditions. Response Time (RT) was averaged for both trial conditions for each participant to draw a comparison between performance in pre-intervention and post-intervention tests. Accuracy (in percentage) for the two same conditions was also calculated using data from ePrime 3.0 based on number of correct trials out of total trials. RT and Accuracy were compared in the two conditions for analysing the effect of meditation intervention on behavioural performance.

### IV. RESULTS

*1) Behavioural results*

Using reaction times and accuracy as variables, behavioral traits were compared between pre- and post-intervention conditions. Preliminary observation from Fig. 3 is that reaction time for incongruent trials is much larger than congruent trials (612ms < 714ms) before intervention. This observation depicts Stroop interference effect (Stroop, 1935) i.e., induced conflict in incongruent trials as a result of which it slows response time. Fig. 3a shows RT differences between trials in the two conditions, in pre-intervention condition, average RT for incongruent trials across all participants is 714ms which reduces to 662ms in post-intervention condition. A major reduction in RT for congruent trials is also observed: in pre-intervention conditions, average RT for congruent trials is 612ms which reduces to 584ms in post-intervention conditions. The accuracy (in percentage) of responses for both trial types in the two conditions is shown in Fig. 3b. The pattern of increment in accuracy of responses in post-intervention trials of both categories viz., congruent and incongruent represent the positive impact of intervention on efficiency of attentional networks while resolving conflict in incongruent trials as well as the ability to respond to congruent trials faster and more efficiently.

*2) Event-related potential (ERP) results*

After stimulus presentation, major ERP components were manually studied for average signals of incongruent trials of Stroop task across all subjects. The signals in both conditions i.e., pre-intervention and post-intervention were compared for amplitude of signals during conflict-ridden incongruent trials (Fig. 2). Several attention-related ERP components were observed in stimulus-locked signals:

components P200, N200, and P300 were observed at channels Pz and CPz in the time windows of 50–150 ms, 150–250 ms, and 250–350 ms, respectively. The signals of post-intervention in orange colour and pre-intervention conditions are depicted in blue colour. Major amplitude differences can be observed in the aforementioned time windows depicting ERP components P2 and N2 at channels Pz and CPz and P3 components at Pz, CPz, Fz and FCz with significant differences at Fz and FCz electrodes

## V. Discussion

In this intervention study, a classic attention network task was implemented to study the effect of a brief meditation session on attention processes. By inducing conflict, we focused on observing the direct implication of intervention on attentional capacity. The findings of this study indicate positive effects on the efficiency of participants in recognizing and responding to conflicting stimuli.

Past studies [44], [45] have shown that P200 (around 200ms post-stimulus) relates to automatic attention required for early perceptual processes. Positive deflection in amplitude between 150-250ms is observed as an increase in P200 component, hence depicting increased brain's capacity to allocate more resources in perceptual processing for stimulus evaluation and conflict detection. Another major ERP component that is important during visual processing is N200 which is majorly found to have a positive difference in amplitude after intervention. At channels CPz and Pz, N200 appears in a large proportion. Several studies [38], [46] have found N200 to be a conflict-sensitive component as well as having a role in detecting oddball stimuli.

Previously discussed attentional networks proposed by Posner and Petersen (1990) [47] can be correlated with the present findings. The alerting network is recognized for achieving and maintaining an isolated alert state while attending to a stimulus. This function of the attentional network has been associated with frontal and parietal regions of the right hemisphere of the brain. The ERP component P300 is widely associated with this component of attentional network systems. Therefore, elevation in P300 amplitude in frontal channels indicates efficient attention-related processes.

## VI. Limitation and Future Scope

Throughout the analysis of present study, it has been observed that even brief periods of meditation sessions have led to significant increase in multiple attributes of attentional capacity of non-meditating participants. However, with a larger sample size, such results would be more reliable. Previous neuropsychological studies from the same lab as the present authors have also explored the effect of long-term meditation on conflict monitoring by using spatial analysis. In the paper [48], authors observed less spread of activation throughout the brain regions involved in conflict-resolution in meditators as compared to controls. This is to represent that brain regions are executed more efficiently in meditators while resolving conflict. This indicates that practising meditation strengthens attentional networks allowing fewer brain resources to act more efficiently and faster than in previous conditions.

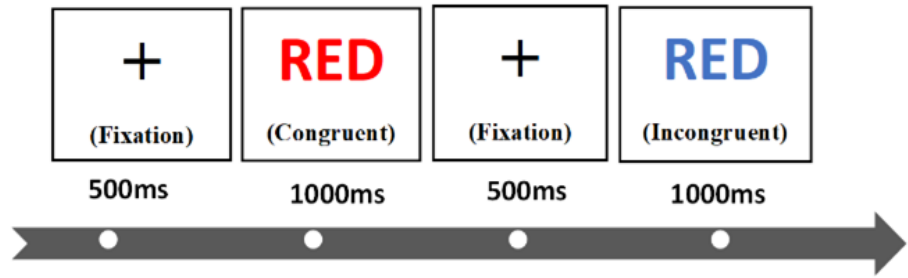

Figure 1: Stroop word-colour experiment design in E-prime 3.0 (Psychology Software Tools, Sharpsburg, PA, USA). The design is composed of congruent stimuli (e.g., word GREEN written in green colour ink) and incongruent stimuli (e.g., word GREEN written in red ink). The stimuli were presented for 1000ms followed by a fixation cross (a 'plus' sign) in black colour on a white screen for 500ms.

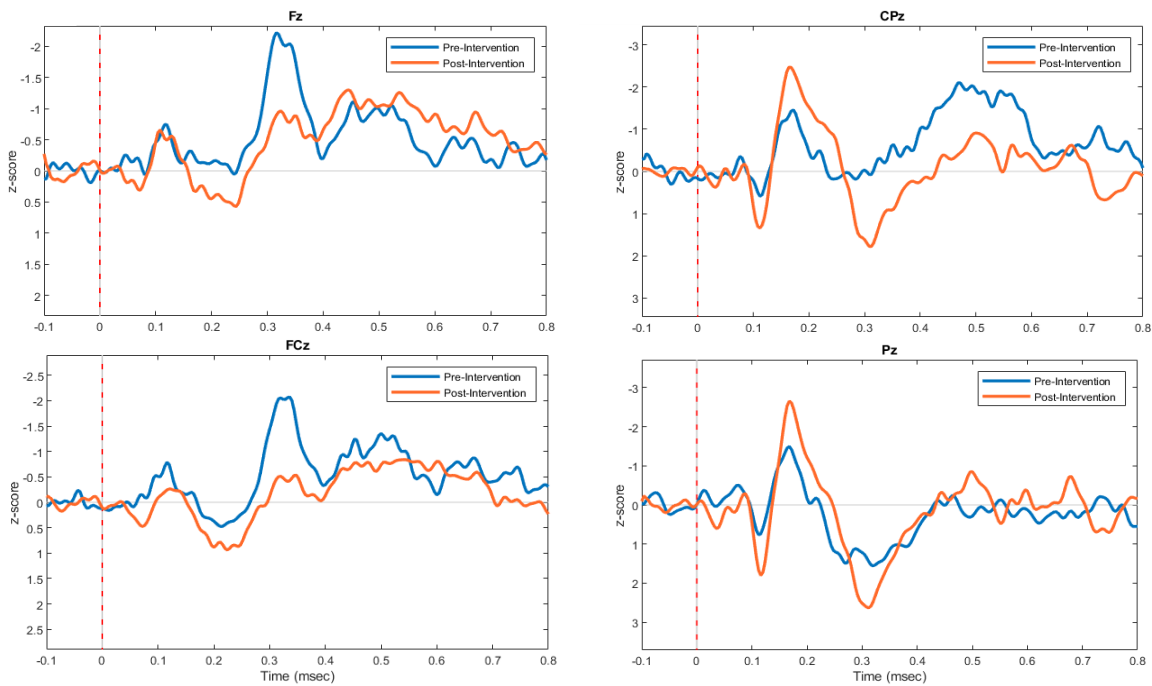

Figure 2. Evoked-response potential (Stimulus-locked analysis) for pooled channels CPz, Pz, FCz, Fz for amplitude comparison between post-intervention (green colour) and pre-intervention (red colour) conditions for cognition-related components P2, N2 and P3 in the time window 50–150 ms, 150–250 ms 250ms-350ms.

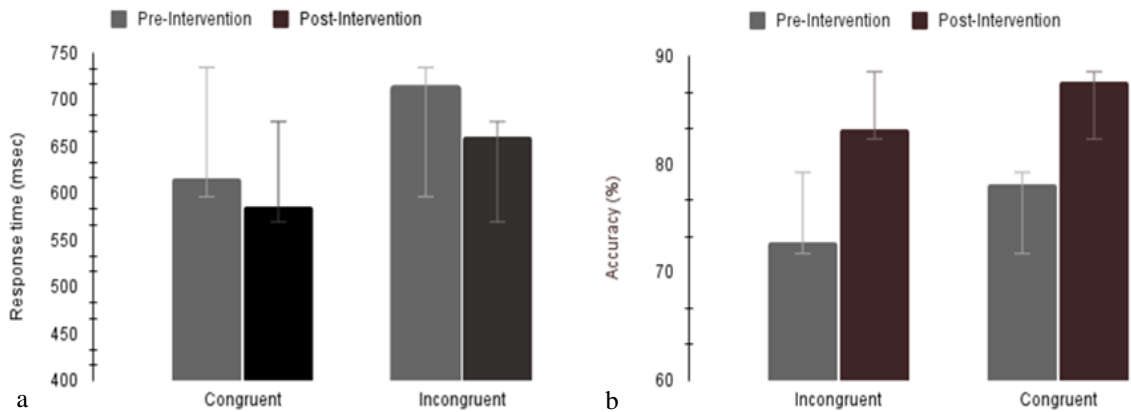

Figure 3. Behavioural analysis: a) Average Response Time (RT) for congruent and incongruent trial conditions in pre-intervention and post-intervention, b) Accuracy (in percentage) for congruent and incongruent trial conditions in pre-intervention and post-intervention. T-test analysis shows that there is a statistically significant difference between the accuracy of both types of trials in pre-intervention and post-intervention condition.